\documentclass[aps,pra,11pt,tightenlines]{revtex4-1}

\usepackage{color}
\usepackage{mathtools}
\usepackage{units}
\usepackage{amssymb}
\usepackage{amsmath}
\usepackage{amsthm}
\usepackage{graphicx} 
\usepackage{bm}
\usepackage{braket}
\usepackage[usenames,dvipsnames,svgnames,table]{xcolor}

\newcommand{\re}{\mathbf{r}}
\newcommand{\pa}{\partial}
\newcommand{\pt}{\partial_t}
\newcommand{\m}[1]{_{\mathrm{#1}}}
\newcommand{\n}[1]{^{\mathrm{#1}}}
\newcommand{\SAE}{\n{SAE}}
\newcommand{\TI}{\n{TICA}}

\newcommand{\A}{\mathbf{A}}
\newcommand{\eu}{\mathrm{e}}

\begin{document}
  
  \title{Many-electron effects of strong-field ionization described in an exact one-electron theory}
  \author{Jakub Koc\'ak}
  \affiliation{ETH Z\"urich,  Laboratorium f\"ur Physikalische Chemie,  8093 Z\"urich, Switzerland}
  \author{Axel Schild}
  \affiliation{ETH Z\"urich,  Laboratorium f\"ur Physikalische Chemie,  8093 Z\"urich, Switzerland}

  \begin{abstract}
    If one-electron observables of a many-electron system are of interest, a many-electron dynamics can be represented exactly by a one-electron dynamics with effective potentials.
    The formalism for this reduction is provided by the Exact Electron Factorization (EEF).
    We study the time-dependent features of the EEF effective potentials for a model of an atom ionized by an ultrastrong and ultrashort laser pulse, with the aim of understanding what is needed to develop computationally feasible approximations.
    It is found that the simplest approximation, the so-called time-independent conditional amplitude (TICA) approximation, is complementary to single-active electron (SAE) approaches as it reproduced the exact dynamics well for high photon frequencies of the laser field or large Keldysh parameter.
    For relatively low frequencies of the laser field or for smaller Keldysh parameters, we find that excited state dynamics in the core region of the atom leads to a time-dependent ionization barrier in the EEF potential.
    The time-dependence of the barrier needs to be described accurately to correctly model many-electron effects, and we conclude that a multi-state extension of the TICA approximation is a possible route how this can be achieved.
    In general, our study sheds a different light on one-electron pictures of strong-field ionization and shows that many-electron effects for such processes may be included by solving a one-electron Schr\"odinger equation, provided the core dynamics can be modeled successfully.
  \end{abstract}
  
  \maketitle
  
  \section{Introduction}
  
    Ultrashort light pulses with a duration of only a few attoseconds provide direct access to the quantum dynamics of electrons in atoms, molecules, and bulk systems \cite{pazourek2015,calegari2016,nisoli2017,biswas2020}.
    Thus, attosecond spectroscopy promises unprecedented possibilities for testing fundamental concepts of chemistry, like electronic structure principles or reaction mechanisms.
    However, the necessary theoretical modeling of molecules interacting with strong and short light pulses is challenging, in particular because both bound electrons and ionized electrons need to be described accurately \cite{palacios2019}.
    This situation leads to new technical and method developments in the area of quantum dynamics, e.g.\ to testing the applicability of time-dependent Density Functional Theory (DFT) \cite{bruner2017,sato2018}, explicit coupling of bound and continuum states \cite{palacios2019}, or first steps towards one-dimensional many-electron models \cite{majorosi2018,majorosi2020}.
    
    Despite these developments, analytical and numerical approaches to electron dynamics in strong laser fields often rely on a Single-Active Electron (SAE) assumption \cite{Schafer1993,Yang1993,Walker1994,Awasthi2008,Ivanov2014} where the dynamics of one electron in some effective potential is considered.
    The underlying idea is that only one ``active'' electron is mainly influenced by the laser field and the other electrons are treated as ``frozen''. 
    By making the SAE assumption, a time-dependent Schr\"odinger equation (TDSE) for one electron in a (classical) laser field is solved and, in this way, many experimental findings can be explained qualitatively.
    However, the SAE assumption has challenges.
    For example, finding suitable effective one-electron potentials is an obstacle, especially for systems with more than one nucleus.
    In such systems the potential is not spherically symmetric and educated guesses based e.g.\ on the density may be useful \cite{abu-samha2010}.
    Additionally, the SAE assumption is an assumption and not an approximation in the sense that, to our knowledge, there exists no procedure which yields an SAE picture as a limit and which can systematically be improved towards the exact result.
    It is also sometimes implied that the SAE assumption does not allow to treat many-electron effects \cite{gordon2006,Ishikawa2015}, although recent studies suggest that e.g.\ field screening effects due to polarization of the other electrons can be included in an SAE approach by hand \cite{romanov2020,abu-samha2020}.
    
    To clarify, the SAE assumption does include many-electron effects via the effective potential.
    It does not, however, describe dynamic changes of the effective many-electron interaction as they may occur e.g.\ during an ionization process.
    Notwithstanding this, a one-electron theory can incorporate all many-electron effects in principle exactly via time-dependent potentials.
    In particular, a many-electron description can be reduced to a one-electron description when one-electron observables are of interest.
    Then, the observables may be obtained in a straightforward way via the one-electron wavefunction obtained from a one-electron Schr\"odinger equation.
    Effects like interaction with the laser field, with nuclei, and with other electrons, are then part of effective one-electron potentials and neither photons, nor nuclei, nor other electrons need to be included in the quantum description explicitly as particles.
    
    A reduction of a quantum system of, say, $n$ particles, to a quantum system of $m < n$ particles, is typically based on a semi-classical approximation, e.g.\ when an electron-laser interaction is modeled with a classical laser field or when nuclei are treated as classical particles in the Born-Oppenheimer approximation.
    However, such a reduction of a quantum system can be made without making approximations by using the Exact Factorization method \cite{abedi2010,abedi2012,gonze2018}:
    The $n$-particle probability density $|\psi|^2$ is written as product of a marginal $m$-particle probability density $|\chi|^2$ and a conditional $(n-m)$-particle probability density $|\phi|^2$.
    The wavefunction $\phi$ describes the $(n-m)$-particle subsystem but also depends parametrically on the remaining $m$ particles.
    $\phi$ can be used e.g.\ to include quantum effects of the nuclei in a Born-Oppenheimer-like treatment of electrons in a molecule \cite{agostini2018}, or to understand why time is a parameter in quantum mechanics \cite{briggs2000,schild2018}, and the Exact Factorization can naturally be applied multiple times up to a point where only single-particle wavefunctions $\phi_1(1), \phi_2(2;1), \phi_3(3;1,2), \dots$ are left that depend successively on more and more parameters \cite{cederbaum2015}.
    In contrast, $\chi$ represents the full $n$-particle system but in terms of only $m$ particles, with the effect of the remaining $(n-m)$-particles included as scalar and/or vector potentials. 
    For instance, $\chi$ can represent the dynamics of a molecule in terms of a nuclear wavefunction $\chi$ alone, where the effect of the electrons is contained in potentials.
    A more abstract use of $\chi$, which considers the theoretical treatment of many electron systems, is that it can represent an electronic wavefunction via a set of spin orbitals (a ``fragment'') embedded in an environment of other spin orbitals \cite{lacombe2020}.
    
    Another interesting case is if $\psi(1,\dots,n)$ is an $n$-electron wavefunction and we choose 
    \begin{align}
      |\psi(1,\dots,n)|^2 = |\chi(1)|^2 \, |\phi(2,\dots,n;1)|^2.
      \label{eq:eef_ansatz}
    \end{align}
    Then $|\chi(1)|^2$ is the one-electron density and $\chi(1)$ is a one-electron wavefunction that is obtained as the solution of a one-electron TDSE.
    It represents the whole $n$-electron system, because it yields (together with the effective one-electron potentials) the one-electron observables of the $n$-electron system, e.g.\ the expectation value of the position or momentum operator.
    One of the authors introduced \eqref{eq:eef_ansatz} as Exact Electron Factorization (EEF) \cite{schild2017}, but the idea was already introduced some time before \cite{hunter1986,hunter1987} and is also closely related to Orbital-free DFT \cite{kraisler2020}.
    A related static approach to tunnel ionization inspired by the Born-Oppenheimer approximation was also proposed \cite{brabec2005,zhao2007}.
    The EEF extends previous developments by providing equations to calculate the effective one-electron potentials and by applying the formalism to time-dependent processes, in particular to the electron dynamics in strong ultrashort laser fields.
    Thus, the effective one-electron potentials in the EEF are time-dependent.
    The main topic of the article at hand is the question of how the time-dependent many-electron effects are encoded in the exact effective potentials, as a step towards the ultimate aim of reproducing those effects approximately but efficiently.
    
    In the following, we first describe the EEF in section \ref{sec:eef}.
    Finding the exact one-electron EEF potentials seems to be at least as hard as solving the full problem.
    Hence, we want to identify the relevant features of the exact one-electron potentials for different laser field parameters and the origin of these features, with the aim of learning what needs to be approximated.
    To achieve this, we consider a simple (spinless) two-electron model of an atom in one dimension, because it already shows relevant many-electron effects but can also be solved numerically for a variety of field parameters.
    The system is presented in section \ref{sec:model}, which is followed in section \ref{sec:sae} by a conceptual comparison to an SAE assumption based on Kohn-Sham (KS) DFT.
    In section \ref{sec:tdd} the time-dependent behavior of the effective one-electron potentials is presented and analyzed. 
    Finally, in section \ref{sec:con} we discuss what still needs to be learned and how the path towards the simulation of realistic many-electron systems may look like.
    
  \section{The Exact Electron Factorization}
    
    \label{sec:eef}
    
    In non-relativistic quantum mechanics, the wavefunction of a system of $n$ electrons can be written as sum of electron permutations of a product $\psi(1,\dots,n;t) \times \xi(1,\dots,n)$, where $\psi(1,\dots,n;t) = \psi(\re_1,\dots,\re_n;t)$ is a spatial wavefunction that depends on the time parameter $t$, and $\xi(1,\dots,n)$ is a spin wavefunction \cite{shpilkin1996}.
    To simplify the discussion, we write the equations for $n=2$ and we only consider the spatial wavefunction $\psi(\re_1, \re_2;t)$.
    Generalization to $n>2$ is straightforward by considering $\re_2$ to be the coordinates of all but one electron, see the supplemental information of \cite{schild2017}.
    
    In the EEF we write the joint probability density $|\psi(\re_1, \re_2;t)|^2$ as product of a marginal probability density $|\chi (\re_1;t)|^2$ and a conditional probability density $|\phi (\re_2; \re_1, t)|^2$, or
    \begin{equation}
      \psi(\re_1, \re_2;t) = \chi(\re_1;t) \phi(\re_2; \re_1, t),
      \label{eq:eef}
    \end{equation}
    where $\chi(\re_1;t)$ is the marginal amplitude and $\phi(\re_2; \re_1, t)$ is the conditional amplitude.
    Below, $\psi$, $\chi$ and $\phi$ are functions that always depend on $\re_1, \re_2$, and $t$ as indicated in \eqref{eq:eef} and those dependencies are only repeated for emphasis.
    We require the partial normalization condition 
    \begin{equation}
      \braket{\phi(\re_2; \re_1, t)| \phi(\re_2; \re_1, t)}_2 = 1  , ~~~~ \forall \re_1, t 
      \label{eq:norm}
    \end{equation}
    where $\braket{\dots | \dots}_2$ denotes the inner product over electron coordinate(s) $\re_2$.
    If $\psi$ is normalized to the number of electrons, $\Braket{\psi|\psi} = n$, we have that
    \begin{align}
      |\chi(\re_1;t)|^2 = \Braket{\psi(\re_1, \re_2;t)|\psi(\re_1, \re_2;t)}_2
      \label{eq:chi2}
    \end{align}
    is the one-electron density which is also normalized to the total number of electrons, $\Braket{\chi(x_1;t)|\chi(x_1;t)}_1 = n$.
    We note that the magnitude of $\chi$ is determined by \eqref{eq:chi2} but its phase is can be chosen, as discussed below.
    Otherwise, \eqref{eq:eef} and \eqref{eq:norm} define the marginal amplitude $\chi$ and the conditional amplitude $\phi$ unambiguously.
    
    The wavefunction $\psi(\re_1, \re_2;t)$ is determined from the TDSE (we use atomic units throughout the text)
    \begin{align}
      i \pt \psi &= \left(\sum_{j=1}^2 \left(\hat{h}(j) + \mathbf{F}(t) \cdot \re_j \right) + V(\re_1,\re_2) \right) \psi
      \label{eq:tdse}
    \end{align}
    with electron-electron interaction $V(\re_1,\re_2)$ and with $t$-independent one-electron Hamiltonian
    \begin{align}
      \hat{h}(j) = -\frac{\nabla_j^2}{2} + V_{\rm ext}(\re_j),
    \end{align}
    where $V_{\rm ext}$ is the external potential due to the presence of the nuclei.
    We write the interaction with the laser field $\mathbf{F}(t)$ in \eqref{eq:tdse} using the dipole approximation and in the length gauge.
    The EEF formalism extends to more complicated Hamiltonians, e.g.\ such that include a vector potential, but these will not be discussed here.

    The equations of motion for the marginal and conditional amplitude can be derived algebraically or variationally.
    For the marginal amplitude $\chi(\re_1;t)$, the equation of motion is 
    \begin{equation}
      i \pa_t  \chi = \left(\frac{1}{2} \left(- i \nabla_1 + \A(\re_1;t)\right)^2  + \varepsilon(\re_1;t)  \right)  \chi  , \label{eq:tDSE-MA}
    \end{equation}
    with $t$-dependent scalar potential $\varepsilon(\re_1;t)$ and vector potential $ \A(\re_1;t)$. 
    The scalar potential (hereafter called EEF potential) is given by
    \begin{equation}
      \varepsilon(\re_1;t) = V\m{ext}(\re_1) + \varepsilon\m{av} + \varepsilon\m{F} + \varepsilon\m{FS} + \varepsilon\m{GD} \label{eq:eps}
    \end{equation}
    where 
    \begin{equation}
      \varepsilon\m{av}(\re_1;t) = \Braket{\phi|\hat{h}(2) + V(\re_1,\re_2)|\phi}_2
      \label{eq:epsav}
    \end{equation}
    is the average kinetic and potential energy of the electron(s) at $\re_2$ given one electron is clamped at $\re_1$,
    \begin{equation}
      \varepsilon\m{F}(\re_1;t) = \varepsilon\m{F1}(\re_1;t) + \varepsilon\m{F2}(\re_1;t) 
    \end{equation}
    represents interaction with the laser field via the usual one-electron interaction 
    \begin{align}
      \varepsilon\m{F1}(\re_1;t) &= \mathbf{F}(t) \cdot \re_1
    \end{align}
    and an additional interaction 
    \begin{align}
      \varepsilon\m{F2}(\re_1;t) &= \mathbf{F}(t) \cdot \mathbf{d}(\re_1;t)
      \label{eq:epsf2}
    \end{align}
    with a $t$-dependent dipole contribution $\mathbf{d}(\re_1;t) = \braket{\phi| \re_2 |\phi}_2$,
    \begin{equation}
      \varepsilon\m{FS}(\re_1;t) = \frac{1}{2} \braket{\nabla_1 \phi| \left(1 - \ket{\phi} \bra{\phi} \right) | \nabla_1 \phi}_2  ,
      \label{eq:epsfs}
    \end{equation}
    is a geometric term that is needed because the electron at $\re_1$ is actually not clamped and that is related to the Fubini-Study metric \cite{provost1980}, and
    \begin{equation}
      \varepsilon\m{GD}(\re_1;t) = \braket{\phi|-i \pa_t|\phi}_2
      \label{eq:epsgd}
    \end{equation}
    is a gauge-dependent term.
    The (gauge-dependent) vector potential $\A(\re_1;t)$ is
    \begin{equation}
      \A(\re_1;t) = \braket{\phi|-i \nabla_1|\phi}_2.
      \label{eq:epsvp}
    \end{equation}
    All these potentials carry a $t$-dependence because of the $t$-dependent conditional wavefunction $\phi$ and, in this way, encode the $t$-dependent many-electron interaction.
    
    As mentioned above, the phase $\arg\left( \chi(\re_1,t) \right)$ of the marginal amplitude is arbitrary and the transformation $(\chi,\phi) \rightarrow (\widetilde\chi,\widetilde\phi)$ with
    \begin{align}
      \widetilde\chi(\re_1;t)        &= \eu^{-i S(\re_1;t)} \chi(\re_1;t) \\
      \widetilde\phi(\re_2;\re_1, t) &= \eu^{+i S(\re_1;t)} \phi(\re_2;\re_1, t),
    \end{align}
    for real-valued $S(\re_1;t)$ leaves the total wavefunction \eqref{eq:eef} unchanged, fulfills the partial normalization condition \eqref{eq:norm}, and leaves the equations of motion for $\chi$ and $\phi$ (see below) invariant provided the potentials are changed as 
    \begin{align}
      \widetilde\A  &= \A + \nabla_1 S \\
      \widetilde\varepsilon\m{GD}  &= \varepsilon\m{GD} + \pa_t S.
    \end{align}
    Thus, the choice of $S(\re_1;t)$ fixes a gauge.
    
    For the EEF it is important to note that $\chi$ is determined from a one-electron TDSE \eqref{eq:tDSE-MA}, that $\rho(\re_1;t) = |\chi|^2$ is the exact one-electron probability density and $\mathbf{j}(\re_1;t) = \operatorname{Im}\left(\chi^* \nabla_1 \chi \right) + \mathbf{A} |\chi|^2$ is the exact one-electron probability current density.
    Also, the one-electron expectation values for position, momentum, and kinetic energy are given as
    \begin{align}
      \braket{\re_1}(t) &= \frac{1}{N} \Braket{\chi | \re_1 | \chi}_1 \\
      \braket{\mathbf{p}_1}(t) &= \frac{1}{N} \Braket{\chi | -i \nabla_1 + \A | \chi}_1 \\
      \braket{T_1}(t) &= \frac{1}{N} \Braket{\chi | \frac{1}{2} [- i \nabla_1 + \A]^2  + \varepsilon\m{FS} | \chi }_1
    \end{align}
    with $N = \Braket{\chi | \chi}_1$.
    Consequently, the marginal one-electron amplitude $\chi(\re_1;t)$ together with $\mathbf{A}$ and $\varepsilon\m{FS}$ yield essentially all relevant one-electron quantities (operators that contain any power of $\re_1$ as well as the first and second derivative with respect to $\re_1$), but for the total $n$-electron system.
    Thus, we call $\chi$ the EEF wavefunction in the following.
    We note that $(-i \nabla_1 + \A)$ is the canonical momentum operator and that $\mathbf{A}$ and $\varepsilon\m{FS}$ can be combined into the quantum geometric tensor which describes the effect that the presence of the electron(s) at $\re_2$ has on the wavefunction $\chi(\re_1;t)$ for infinitesimal changes of $\re_1$ \cite{berry1989}.
    
    In attoscience, typical observables are the high-harmonic generation spectrum and one-electron ionization rates.
    For the choice of gauge $\A = 0$ (which is only possible iff $\nabla_1 \wedge \A = 0$ \cite{requist16}), those observables can be determined from $\chi$ alone, hence we would only need to solve a one-electron TDSE \eqref{eq:tDSE-MA} to obtain the observables of the many-electron system.
    However, for this purpose we need the $t$-dependent effective potential $\varepsilon(\re_1;t)$, for which we need to know the conditional amplitude $\phi(\re_2;\re_1,t)$.
    
    The equation of motion for the conditional amplitude is 
    \begin{align}
      \left(i \pt + \hat{C} + \varepsilon(\re_1;t) \right) \phi(\re_2;\re_1,t)
        &= \left(\hat{h}(2) + \hat{U} \right) \phi(\re_2;\re_1,t),
        \label{eq:tDSE-CO}
    \end{align}
    which is a generalized TDSE with operators 
    \begin{align}
      \hat{C} &= -\frac{(-i \nabla_1 + \A)\chi}{\chi} \cdot (-i \nabla_1 - \A) \\
      \hat{U} &= \frac{(-i \nabla_1 - \A)^2}{2}.
    \end{align}
    In terms of these operators, the EEF potential is given by the expression
    \begin{align}
      \varepsilon(\re_1;t) = \Braket{\phi|\hat{h}(2) + \hat{U} - i\pt - \hat{C}|\phi}_2.
    \end{align}
    Solving the coupled equations \eqref{eq:tDSE-MA} and \eqref{eq:tDSE-CO} exactly seems harder than to solve the full many-electron problem \eqref{eq:tdse}, in particular because solving \eqref{eq:tDSE-CO} numerically is mathematically challenging.
    However, if there was a way to find the exact scalar potential $\varepsilon(\re_1;t)$ approximately, only the one-electron TDSE \eqref{eq:tDSE-MA} needs to be solved.
    Thus, in the following we want to learn how the exact scalar potential $\varepsilon(\re_1;t)$ behaves during an ionization process in a strong and ultrashort laser field.
    
  \section{Model}
    
    \label{sec:model}
    
    For this purpose, we study a one-dimensional two-electron system similar to those used e.g.\ in \cite{Bauer1997}.
    It is described by the $t$-dependent wavefunction $\psi(x_1, x_2;t)$ obtained as solution of the TDSE
    \begin{equation}
      i \pa_t  \psi = \left(\hat{H} + F(t) (x_1 + x_2)  \right) \psi , \label{eq:model_tdse}
    \end{equation}
    with 
    \begin{equation}
      \hat{H} = \sum_{j=1}^2 \left(  -\frac{\pa_j^2}{2} + V_{\rm ext}(x_j) \right) + V(x_1,x_2),
    \end{equation}
    where we use the soft-Coulomb potentials 
    \begin{align}
      V_{\rm ext}(x) &= -\frac{2}{\sqrt{c_{\rm en}+x^2}} \\
      V(x_1,x_2)     &=  \frac{1}{\sqrt{c_{\rm ee}+(x_1-x_2)^2}}
    \end{align}
    with parameters \unit[$c_{\rm en} = c_{\rm ee} = 0.55$]{$a_0^2$} to describe the interaction of the electrons with one nucleus and the electron-electron interaction, respectively.
    
    We consider ``spinless'' electrons where the spatial wavefunction is antisymmetric, hence we use as initial state the lowest eigenstate $\psi_0$  of
    \begin{align}
      \hat{H} \psi_j = E_j \psi_j,
      \label{eq:hpsi}
    \end{align}
    where we only consider states with correct symmetry property, $\psi_j(x_1,x_2) = -\psi_j(x_2,x_1)$.
    Our model may also be interpreted as a one-dimensional model of a helium atom, then $\psi_0$ corresponds to its lowest triplet state.
    We chose this state because for the symmetric ground state of $\hat{H}$, KS-DFT and the EEF are identical, as there is only one orbital which both electrons share, and the electron interaction effects which we describe below are largely absent.
    There are also electron-interaction effects for spin-paired electrons occupying the same orbital (see e.g.\ \cite{pazourek2012}) which are, however, not the focus of the investigations presented here.
    We find that the qualitative features of the EEF potentials change when the number of orbitals occupied in a Kohn-Sham picture change.
    Hence, our spinless two-electron model contains effects similar to those that occur for a spin-paired three- or four-electron model where (in the KS picture) only two orbitals are occupied, even though neither the number of electrons nor the nuclear charge matches -- as can be seen by comparing the behavior of the model as presented below with the spin-paired three-electron model used in \cite{schild2017}.
    
    \begin{figure}[htbp]
      \begin{center}
        \includegraphics[scale=0.6]{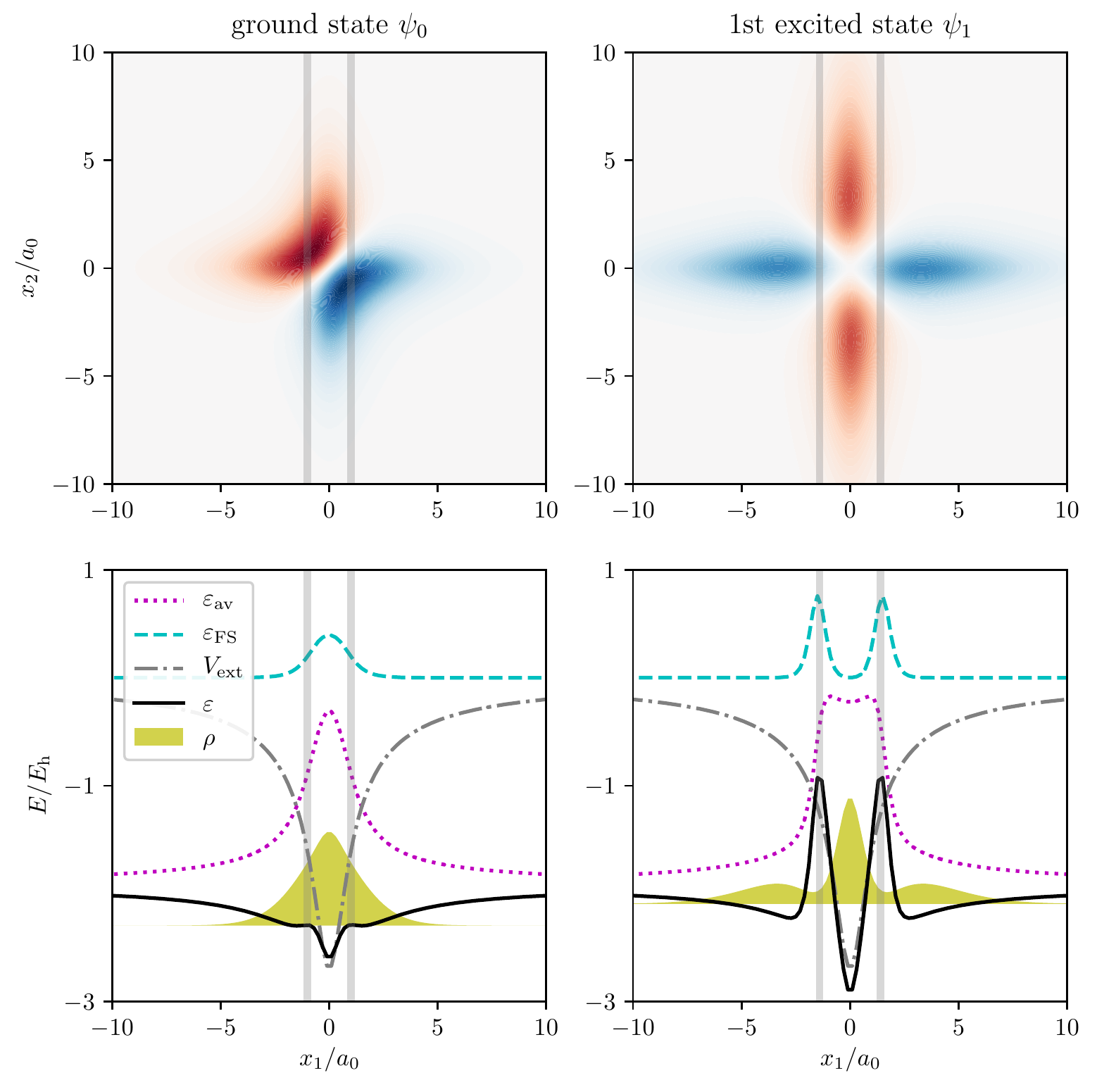}
        \caption{Lowest two anti-symmetric eigenstates $\psi_j$ of \eqref{eq:hpsi} (top) and corresponding EEF potentials $\varepsilon_j$ (bottom). In the bottom panels, also the one-electron density $\rho$ is shown as filled area. Vertical lines indicate the position of maxima of $\varepsilon_j$ which correspond to minima or some depletion of $\rho$.} \label{fig:es}
      \end{center}
    \end{figure}
    
    In Fig.\ \ref{fig:es} the two energetically lowest states $\psi_j$ of \eqref{eq:hpsi} with correct symmetry are shown together with the one-electron densities $\rho_j = |\chi_j|^2$ and potentials $\varepsilon_j$ appearing in the time-independent version of \eqref{eq:tDSE-MA},
    \begin{align}
      E_j \chi_j(x_1) = \left( -\frac{\partial_1^2}{2} + \varepsilon_j(x_1) \right) \chi_j(x_1)
      \label{eq:mtise}
    \end{align}
    with $\varepsilon_j(x_1) = V_{\rm ext}(x_1) + \varepsilon_{\rm av}(x_1) + \varepsilon_{\rm FS}(x_1)$.
    Each state $\psi_j$ corresponds to a reduced potential $\varepsilon_j$ which has $\chi_j$ as its ground state with the energy eigenvalue $E_j$ of the full system.
    The electronic structure is encoded in $\varepsilon_j$, which thus has features like barriers in the core region (see the gray vertical lines in the panels of Fig.\ \ref{fig:es}) that correspond to a suppression of probability density in those regions.
    We note that smaller values of the one-electron density correspond to higher barriers, but that the one-electron density $\rho_j$ and hence $\chi_j$ is never zero.
    The appearance of those barriers (see below) is well known from orbital-free DFT, where the potential $\varepsilon_j(x_1)$ of \eqref{eq:mtise} is to be approximated, typically as functional of the one-electron density \cite{finzel2016}.
    
    We choose the 6-cycle laser pulse 
    \begin{align}
      F(t) = F_0 E_{\rm e}(t) \cos(\omega_0 t) 
    \end{align}
    with envelope function $E_{\rm e}(t)$ that increases quadratically as $(t/t_{\rm on})^2$ during the first two cycles, $t_{\rm on} = 2 \frac{2 \pi}{\omega_0}$, is $1$ during the next two cycles, decreases quadratically during the following two cycles, and is zero otherwise.
    The 6-cycle laser pulse depends on two parameters: 
    The central angular frequency $\omega_0$ and the maximum amplitude of the laser field $F_0$. 
    We consider values of the angular frequency $\omega_0$ ranging from \unit[$0.1$]{$E\m{h} /\hbar$} (wavelength \unit[$456$]{$\mathrm{nm}$}) to \unit[$1.0$]{$E\m{h} /\hbar$} (wavelength \unit[$46$]{$\mathrm{nm}$}) and three different maximal amplitudes of the laser field $F_0$, \unit[$0.015$]{$E\m{h} /(e a_0)$}, \unit[$ 0.030$]{$E\m{h} /(e a_0)$}, and \unit[$0.050$]{$E\m{h} /(e a_0)$}, which correspond to the intensities \unit[$7.9 \times 10^{12}$]{$\mathrm{W/cm^2}$}, \unit[$3.2 \times 10^{13}$]{$\mathrm{W/cm^2}$}, and $8.8 \times 10^{13}  \mathrm{W/cm^2}$, respectively.

\begin{figure}[htbp]
      \begin{center}
        \includegraphics[scale=0.6]{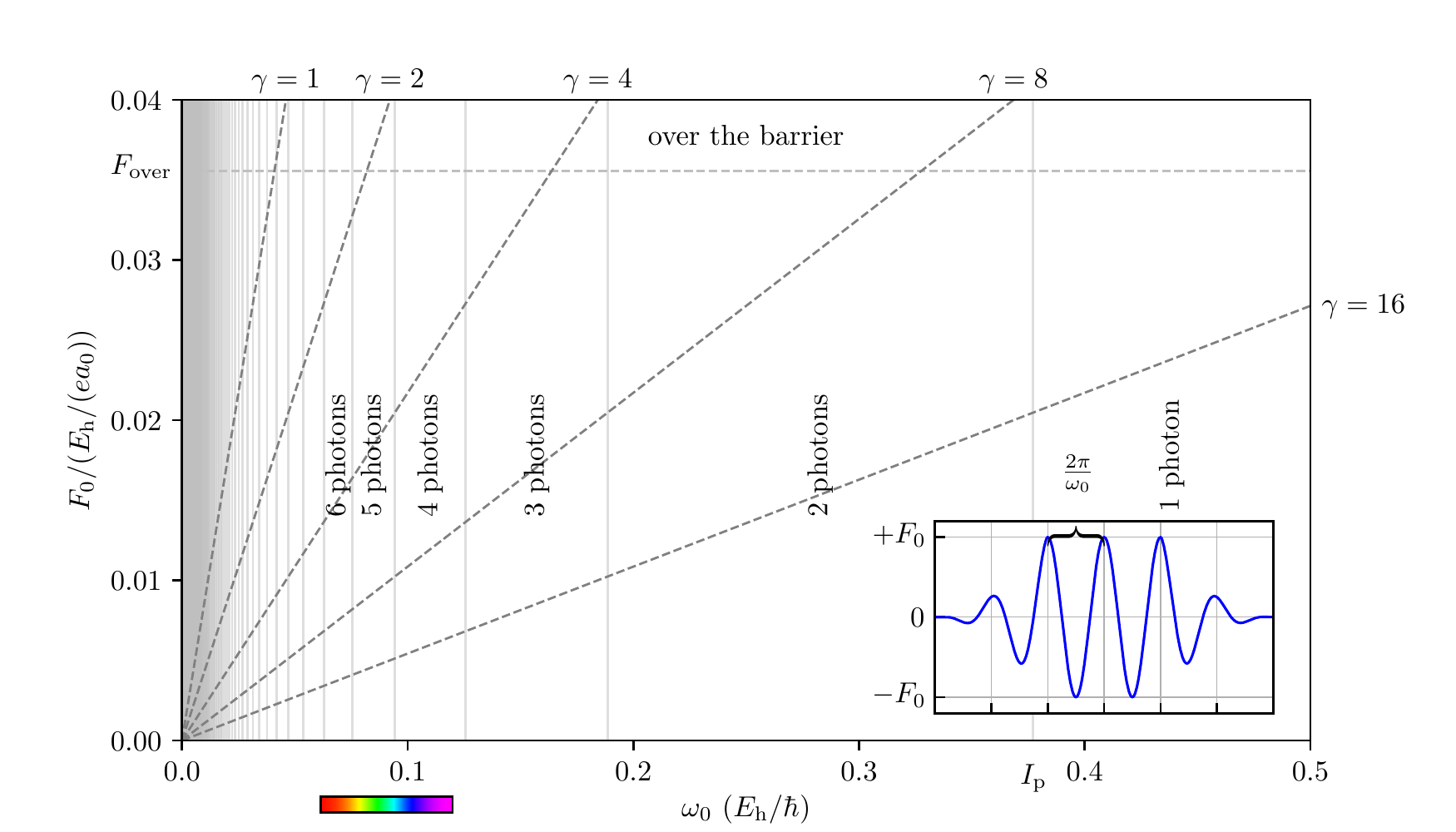}
        \caption{
          Parameter space of the 6-cycle laser pulse used in our simulations with a central angular frequency $\omega_0$ and maximal amplitude of the laser field $F_0$ for our model with an ionization potential \unit[$I\m{p} = 0.377$]{$E\m{h}$}. 
          We mark the necessary electric field strength $F\m{over}$ for over-the-barrier ionization, for $Z=1$, as well as regions of different Keldysh parameters $\gamma$ and the minimal number of absorbed photons required for ionization, given by $\lceil I\m{p} / \omega_0 \rceil$. 
          The colorbar below the abscissa indicates the position of the visible part of the spectrum.} \label{fig:para}
      \end{center}
    \end{figure}

    There are qualitatively different regimes depending on the field parameters $\omega_0$ and $F_0$, as well as on the Keldysh parameter $\gamma = \sqrt{2 I\m{p}} \omega_0 / F_0$ \cite{Keldysh1965}.
    The Keldysh parameter combines the $I\m{p}$ as relevant system parameter with the field parameters $\omega_0$ and $F_0$.
    One regime defined by $F_0$ is over-the-barrier ionization, where the field strength is strong enough that electron(s) can freely leave the core region without the necessity of tunneling. 
    Assuming an asymptotic Coulomb potential $-Z/|x|$, the field strength needs to be larger than $F\m{over} = I\m{p}^2/(4 Z)$ for over-the-barrier ionization \cite{kiyan1991}, with $Z$ being the nuclear charge and $I\m{p}$ being the ionization potential.
    For our model $I\m{p} = 0.377  E\m{h}$, hence calculations with the field strength \unit[$F_0 = 0.050$]{$E\m{h} /(e a_0)$} correspond to over-the-barrier ionization.
    A second regime is tunnel ionization, which happens for $\gamma < 1$ or $\ll 1$ and $F_0 < F\m{over}$, as the electron has enough time to tunnel through barrier within a laser cycle.
    Then, the parameter space of the laser field can also be separated based on the minimum number of absorbed photons required for ionization given by $\lceil I\m{p} / \omega_0 \rceil$:
    For $\hbar \omega_0 > I_{\rm p}$, we have single-photon ionization, while otherwise multi-photon ionization takes place.
    The parameter space of the laser field for our model is shown in Fig.\ \ref{fig:para}.
    A more detailed discussion of the regimes can e.g.\ be found in \cite{amini2019}.
    
    For all numerical eigenstate calculations and time propagations we use QMstunfti \cite{qmstunfti}, which is a Python toolbox designed to solve grid-based quantum mechanics.
    In particular, we use a sparse-matrix representation of the respective Hamiltonian where derivatives are obtained within a finite difference approximation.
    Both for the eigenstate calculation and for the propagation we rely on functionalities of the scipy.sparse module \cite{scipy} which partially uses the ARPACK library \cite{arpack}.
    We use a grid spacing of $0.01 / \omega_0$ for the $t$-grid and of \unit[$0.2$]{$a_0$} for the spatial grid with \unit[$|x_j| < 100$]{$a_0$}.
    To avoid reflections at the grid boundaries, we absorb the wavefunction in the region $90 \, a_0 < |x_j| < 100 \, a_0$ by multiplication with a mask function being $1$ at \unit[$|x_j| = 90$]{$a_0$} and decreasing to 0 until \unit[$|x_j| = 100$]{$a_0$} as $\cos^{1/8}$.
  
  \section{Comparison to a Single-Active Electron assumption}
      
      \label{sec:sae}
      
      A standard approach to attoscience modeling is the SAE assumption, but there are different ways how this assumption can be implemented.
      Here, we consider an SAE model based on KS-DFT with the exact KS-potential.
      
      In KS-DFT, the one-electron density for a spinless $n$-electron system is obtained as 
      \begin{align}
        \rho(\re_1) = \sum_{j=0}^{n-1} |\varphi_j^{\rm KS}(\re_1)|^2
        \label{eq:ksdft}
      \end{align}
      where $\varphi_j^{\rm KS}(\re_1)$ are the KS-orbitals that are eigenstates of a one-electron Hamiltonian with KS-potential $V^{\rm KS}(\re_1)$,
      \begin{align}
        \left( -\frac{\nabla_1^2}{2} + V^{\rm KS}(\re_1) \right) \varphi_j^{\rm KS}(\re_1) = \varepsilon_j^{\rm KS} \varphi_j^{\rm KS}(\re_1).
      \end{align}
      An SAE approach can be defined by
      \begin{equation}
        i \pa_t  \chi\SAE(\re_1;t) = \left(- \frac{1}{2} \nabla_1^2  + V^{\rm KS}(\re_1) + \mathbf{F}(t) \cdot \re_1  \right)  \chi\SAE(\re_1;t)
        \label{eq:tdse_sae}
      \end{equation}
      with initial state $\chi\SAE(\re_1;t=0) = \varphi_j^{\rm KS}(\re_1)$ which is one of the KS-orbitals, typically the highest occupied KS-orbital.
      Thus, in this SAE approach only one orbital is propagated while the others are kept frozen, and we assume that $V^{\rm KS}(x_1)$ is $t$-independent.
      
      \begin{figure}[htbp]
        \begin{center}
        \includegraphics[width=0.9\textwidth]{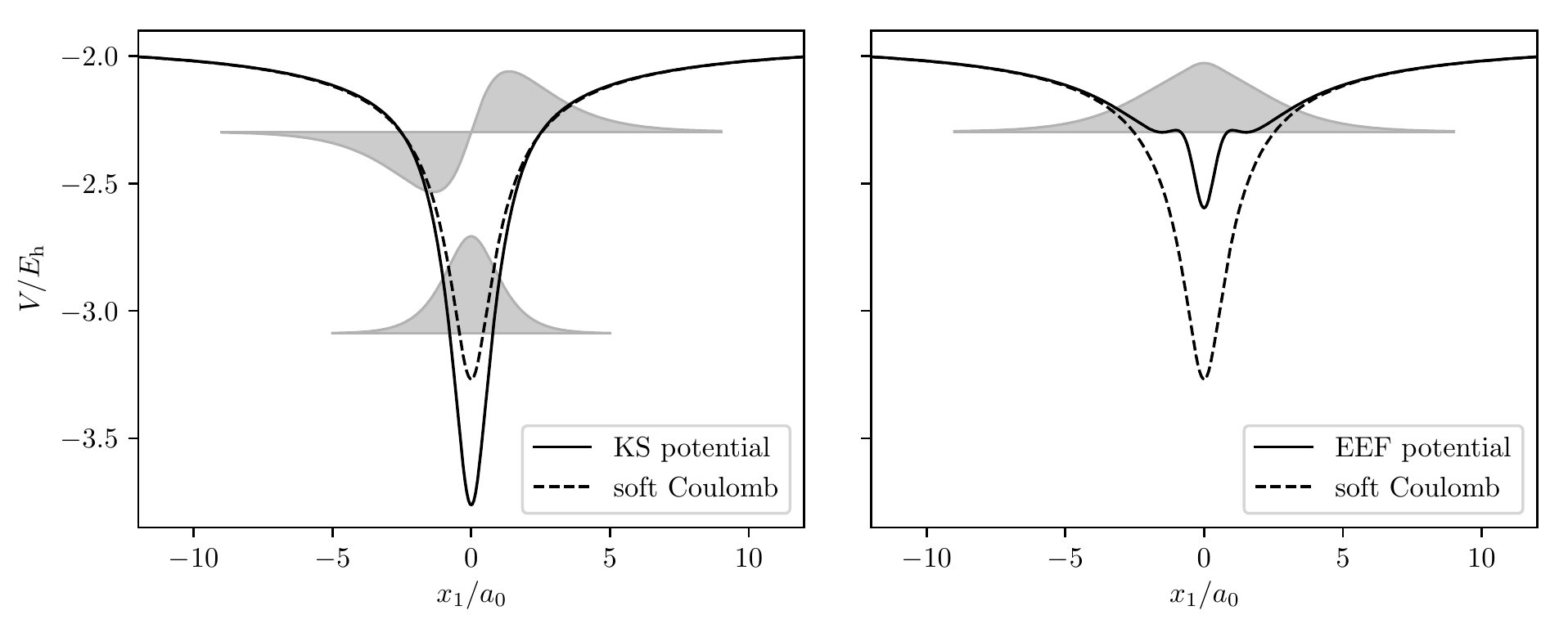}
        \caption{
          Kohn-Sham potential and lowest two Kohn-Sham orbitals (left) as well as the Exact Electron Factorization potential and wavefunction (right) for the antisymmetric initial state of the considered two-electron model. For comparison, the external soft-Coulomb potential $V_{\rm ext}$ is also shown.
          } \label{fig:pic_KSvsEEF}
        \end{center}
      \end{figure}
      
      The presented SAE approach is close to the idea of reconstructing effective SAE potentials for molecules from the static KS potential \cite{Awasthi2008} and it uses the correct ionization potential, which is considered to be a decisive parameter in the SAE assumption \cite{Hofmann2014}.
    
      For our one-dimensional model, the exact KS-potential and KS-orbitals are shown in Fig.\ \ref{fig:pic_KSvsEEF} together with the EEF quantities for the ground state.
      While the EEF describes all electrons with a single one-electron wavefunction $\chi$, KS-DFT relies on multiple orbitals.
      In some sense this an advantage, as KS-DFT maps the interacting many-electron problem to a non-interacting many-electron problem with a wavefunction that is a Slater determinant of the KS-orbitals, and thus has the (anti-)symmetry requirements already contained in the ansatz.
      In contrast, while the product of $\chi$ and $\phi$ is the exact many-electron wavefunction and hence fulfills all relevant symmetry constraints, the ansatz \eqref{eq:eef} does not include the symmetry requirements explicitly.
      Thus, the effective one-electron potential $\varepsilon$ of the EEF is very different compared to the KS-potential. 
      In particular, while the KS-potential looks qualitatively like the soft-Coulomb potential, the EEF potential for the ground state has an additional local barrier at ca.\ \unit[1]{$a_0$}.
      This barrier reflects the electronic structure, but in a way that is somewhat less intuitive than the multi-orbital picture of KS-DFT.
      However, we note that similar barriers can also appear in the KS potential, e.g.\ for an excited symmetric state of a model similar to the model used here \cite{elliott2012b}.
      
      For Fig.\ \ref{fig:pic_KSvsEEF} the potentials are shifted such that the asymptotic energy for $|x_1| \rightarrow \infty$ corresponds to the energy of the cation.
      Thus, the KS-eigenvalue $\varepsilon_1^{\rm KS}$ of the highest occupied KS-orbital is equal to the energy of the two-electron system, which is also the EEF eigenvalue for $\chi$ in the absence of the laser field.
      
      In \cite{schild2017}, a first approximation to the EEF was proposed which looks very similar to the SAE approach and which is computationally feasible for realistic many-electron systems.
      In the time-independent conditional amplitude (TICA) approximation it is assumed that the conditional amplitude does not change during the interaction with the laser field, i.e, $\phi(\re_2; \re_1, t) \approx \phi_0 (\re_2;\re_1)$ for all times $t$.
      When we chose the gauge such that the vector potential is zero, $\A(\re_1;t) = 0$, the TICA Schr\"odinger equation is
      \begin{equation}
      i \pa_t  \chi\TI(\re_1;t) = \left(- \frac{1}{2} \nabla_1^2  + \varepsilon\TI(\re_1) + \mathbf{F}(t) \cdot \left(\re_1 + \mathbf{d}_0(\re_1)\right)  \right)  \chi\TI(\re_1;t)  , \label{eq:TICA}
      \end{equation}
      with time-independent potential $ \varepsilon\TI(\re_1)$ that can be obtained from $\phi_0(\re_2;\re_1)$ or from the initial electron density $\rho_0(\re_1) = |\chi\TI(\re_1;0)|^2$, up to a constant, as
      \begin{align}
        \varepsilon\TI(\re_1) = \frac{\nabla_1^2 \sqrt{\rho_0}}{2 \sqrt{\rho_0}},
      \end{align}
      assuming the initial phase of $\chi\TI$ is zero.
      The dipole operator $d_0(\re_1)$ is given by
      \begin{equation}
        d_0(\re_1) = \braket{\phi_0| \re_2 |\phi_0}_2.
      \end{equation}
      If we compare the TDSE for the TICA approximation \eqref{eq:TICA} and the TDSE for the SAE assumption \eqref{eq:tdse_sae}, we see that both are one-electron approaches with a time-independent effective potential that models the many-electron dynamics.
      However, the initial states and the effective potentials are very different, and the SAE approach models only one electron while the TICA in principle models all electrons.
      Thus, we can expect that their applications are rather different.
      
    \section{Time-dependent dynamics}
      
      \label{sec:tdd}
      
      In the following, we discuss results for the spinless one-dimensional two-electron model.
      We choose the gauge where the vector potential is zero, $A(x_1;t) = 0$, and we calculate all quantities from the solution $\psi(x_1,x_2;t)$ of the two-electron problem for the different laser field parameters.
      Thus, we have both the EEF wavefunction $\chi$ as well as the EEF potential $\varepsilon$ and can compare how features of one of these functions manifest in the other function.
      
      \begin{figure}[htbp]
        \begin{center}
        \includegraphics[width=0.5\textwidth]{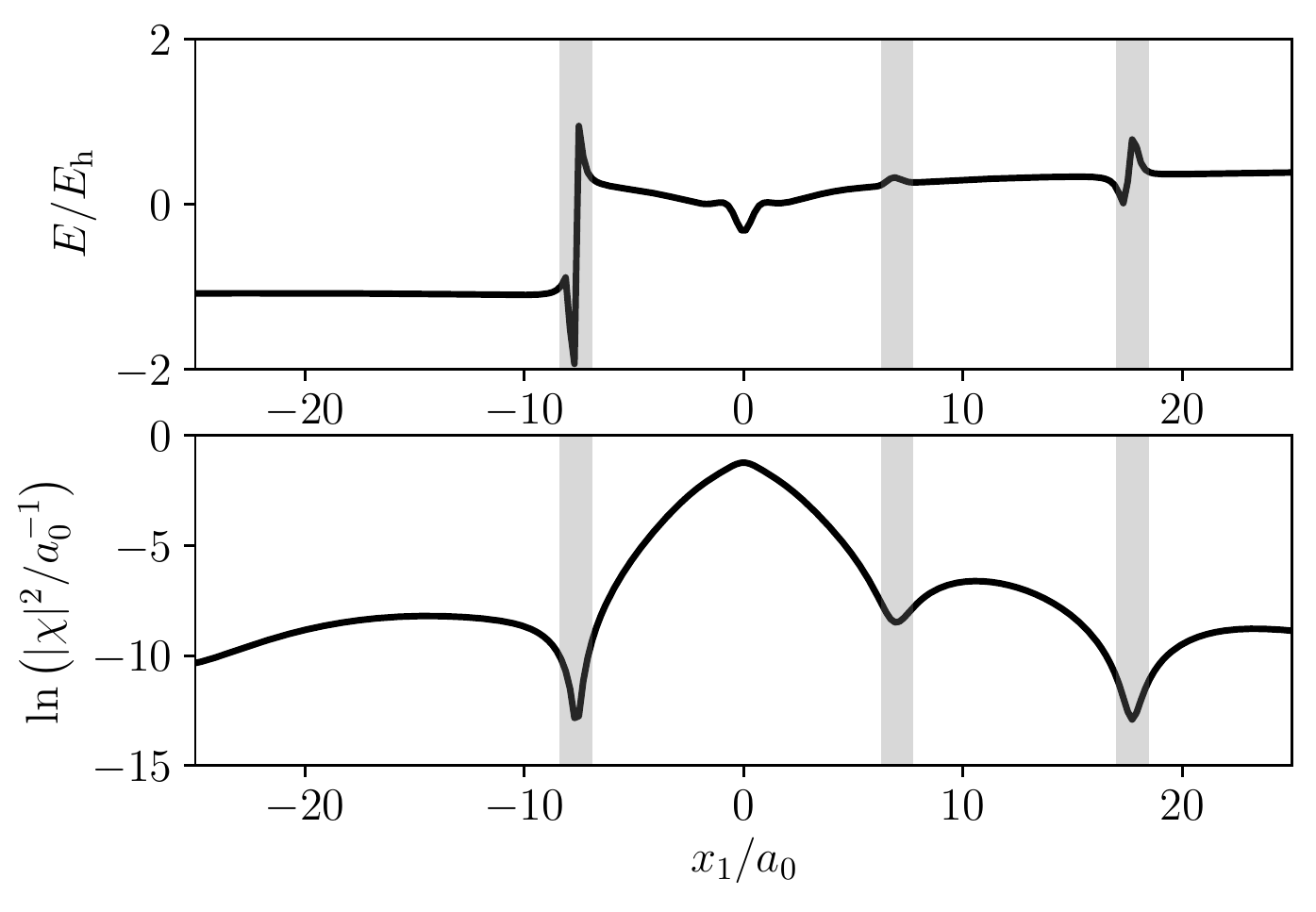}
        \caption{Representative snapshot of the exact one-electron potential $\varepsilon$ (top) and the corresponding one-electron density $\rho = |\chi|^2$ (bottom, shown logarithmically) at some time during the interaction with the laser pulse. Vertical lines indicate the presence of spikes and steps that appear at minima of the density.} \label{fig:pic_sss}
        \end{center}
      \end{figure}
      
      Prominent time-dependent features of the EEF potential are spikes and steps outside the core region, illustrated in Fig.\ \ref{fig:pic_sss}.
      Those spikes and steps appear for some parameters $t$ but also quickly disappear, and it seems that they are rather unimportant features for the construction of suitable approximations.
      
      Spikes typically appear in at the same place and time in the components of the EEF scalar potential, $\varepsilon_{\rm av}$, $\varepsilon_{\rm FS}$, $\varepsilon_{\rm GD}$, $\varepsilon_{\rm F2}$, and the EEF vector potential $\A$.
      From the mathematical formalism of the EEF, the origin of the spikes can be understood by writing \eqref{eq:epsav}, \eqref{eq:epsf2}, \eqref{eq:epsfs}, \eqref{eq:epsgd}, and \eqref{eq:epsvp} in terms of $\psi$ and $\chi$, because for each term we find that it is proportional to $1/|\chi|^2$.
      Alternatively, the spikes can be analyzed by looking at a feature of the EEF wavefunction $\chi$:
      This function can be written in polar representation as 
      \begin{align}
        \chi(x_1;t) = e^{i \theta(x_1;t)} \sqrt{\rho(x_1;t)}
      \end{align}
      with phase $\theta(x_1;t) \in \mathbb{R}$ determined by the choice of gauge and with one-electron density 
      \begin{align}
        \rho(x_1;t) = \Braket{\psi(x_1,x_2;t)|\psi(x_1,x_2;t)}_2.
      \end{align}
      For $|\chi|$ to be zero, we need that $|\psi(x_{1,0},x_2;t_0)| = 0$ for all $x_2$ at some $x_{1,0}$ and some $t_0$.
      While $\psi$ clearly may have nodes, we find that they never lie exactly along a line at some $x_{1,0}$ in $x_2$-direction -- a wavefunctions with this property can be obtained as eigenstate from suitably designed potential, but that such an exactly ``vertical'' node appears during a time-dependent simulation is extremely unlikely.
      Hence, the magnitude $|\chi|$ never reaches zero, but it may become very small.
      However, while such a node is very unlikely from the perspective of the full wavefunction $\psi$, a propagation of $\chi$ by solving the TDSE \eqref{eq:tDSE-MA} for some potential allows in principle for nodes in $\chi$, i.e., for $|\chi|$ to become exactly zero.
      The appearance of nodes is a very common situation when a wavefunction is propagated in some static potential.
      In contrast, the EEF potential $\varepsilon$ has time-dependent spikes of finite height in regions and at times where $|\chi|$ becomes small.
      Scattering at these spikes changes (the phase of) $\chi$ such that the sign change is avoided.
      For an approximate simulation we find that we can ignore the spikes and simply allow the one-electron wavefunction to have nodes, thus these spikes are of little relevance.
      
      The steps appear in the gauge-dependent potential and are equivalent to spikes in the vector potential if a different gauge was chosen.
      From the simulation, we have the vector potential for the gauge $\tilde{\chi} = \sqrt{\rho}$ ($\chi$ being real-valued) given by 
      \begin{align}
        \tilde{A}(x_1;t) 
          = \frac{1}{\rho} \Braket{\psi| -i \pa_1 \psi}_2,
        \label{eq:ag}
      \end{align}
      and we determine the phase $\theta$ of $\chi$ for the gauge $A \stackrel{!}{=} 0$ from 
      \begin{align}
        \theta(x_1;t) = -\int_{-\infty}^{x_1} \tilde{A}(x';t) dx'
        \label{eq:gt}
      \end{align}
      such that $A(x) = \tilde{A}(x') + \pa_1 \theta(x;t) \equiv 0$.
      When $\tilde{A}$ has a spike, we get from \eqref{eq:gt} that the phase $\theta(x_1;t)$ has a step which transfers to the gauge-dependent potential via $\varepsilon_{\rm GD} = \tilde{\varepsilon}_{\rm GD}+ \pt \theta$.
      The steps seems to be related to steps found in DFT \cite{elliott2012,hodgson2017,kraisler2020b} and hint at some qualitative change in the behavior of the electron density $\rho$, but we do not yet have a clear understanding of the steps within the framework of the EEF.
      In DFT, the steps are for example relevant to correctly describe charge transfer and are related to ionization phenomena \cite{lein2005}, hence they might be important in some situations.
      However, as \eqref{eq:ag} suggests we find that the steps in $\varepsilon_{\rm GD}$ always appear where $\rho$ is small (as $\tilde{A} \propto \frac{1}{\rho}$) and we find that they can be ignored for the considered simulations.
      
      To describe an ionization dynamics, a more important many-electron effect encoded in the effective potential is the time-dependence of the core region.
      When we compared DFT with the EEF in Fig.\ \ref{fig:pic_KSvsEEF} above, we noted that the electronic structure of the ground state translates to the EEF as an additional barrier at ca.\ \unit[$|x_1| = 1.0$]{$a_0$}.
      During interaction with the laser field, this barrier can change significantly in height and width.
      Also, the depth of the potential well in the core region may vary.
      
      \begin{figure}[htbp]
        \begin{center}
        \includegraphics[width=0.9\textwidth]{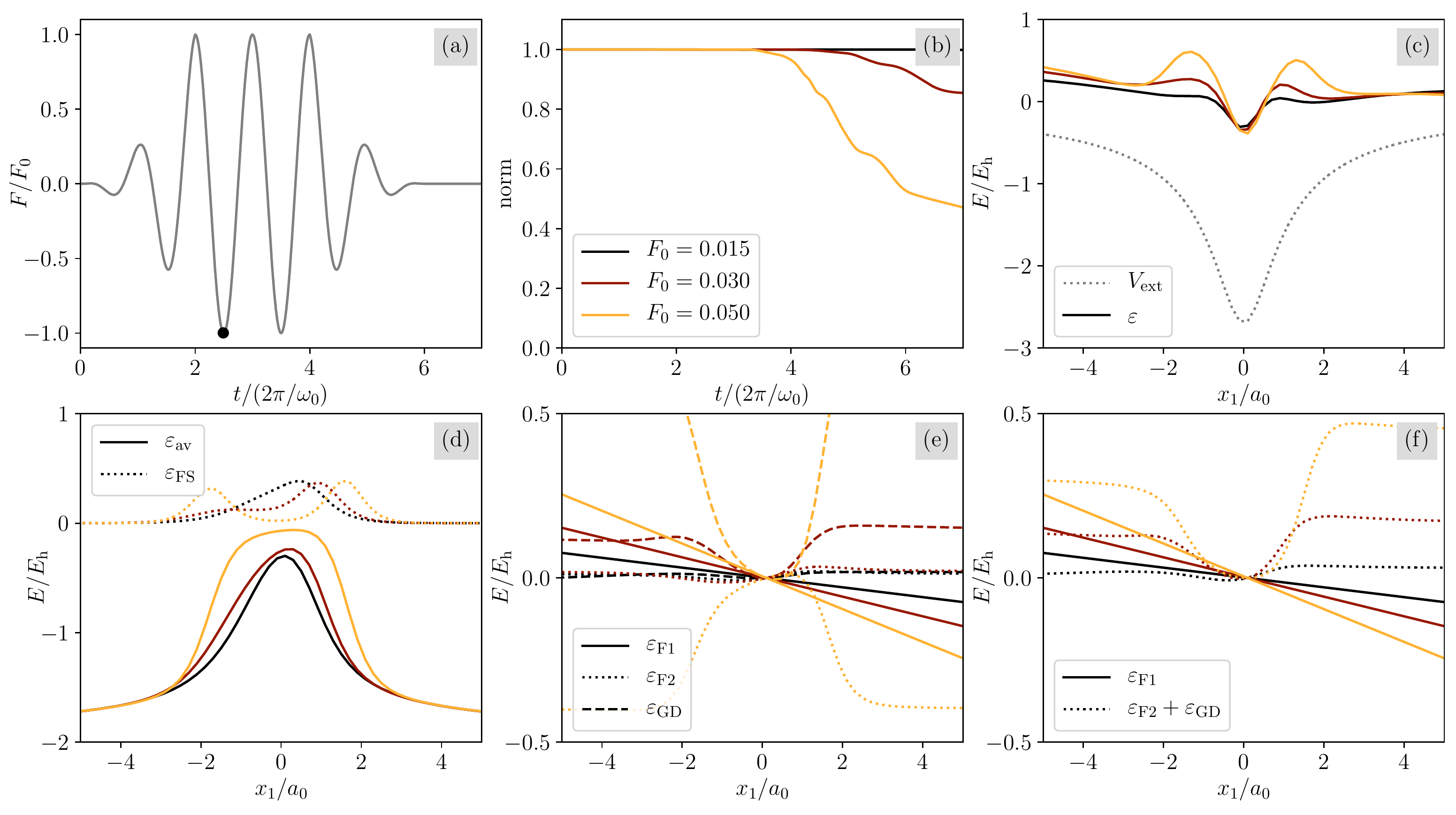}
        \caption{
          (a) 6-cycle laser pulse used in the simulations. The black dot indicates the $t$-parameter at which the potentials in the other panels are shown.
          The other panels are, for laser frequency \unit[$\omega_0=0.1$]{$E_{\rm h}/\hbar$} and for different values of the field strength $F_0$, the norm of the wavefunction (b), the one-electron potential $\varepsilon$ (c) as well as its contributions $\varepsilon_{\rm av}$, $\varepsilon_{\rm FS}$ (d) and $\varepsilon_{\rm F1}$, $\varepsilon_{\rm F2}$, $\varepsilon_{\rm GD}$ (e), (f).} \label{fig:potc1}
        \end{center}
      \end{figure}
      
      In Fig.\ \ref{fig:potc1} we compare different parts of the EEF potential for a laser frequency \unit[$\omega_0 = 0.1$]{$E_{\rm h}/\hbar$} at an instant of time where the laser field amplitude is maximal and the effects in the exact potential are most pronounced.
      Panel (a) of Fig.\ \ref{fig:potc1} shows the laser pulse and the time at which the potentials in the other panels are depicted, while panel (c) shows the exact potential $\varepsilon$ for the three different $F_0$ and in comparison to the external one-electron potential $V_{\rm ext}$.
      The larger $F_0$, the higher is the effective barrier.
      This barrier keeps the electrons bound:
      In panel (b) of Fig. \ref{fig:potc1} the norm of the wavefunction for these cases is shown, which indicates that there is significant ionization happening for the higher field strengths.
      If we think e.g.\ in the SAE KS picture about such an ionization, we expect the electron from the higher orbital to be ionized more easily compared to that in the lower orbital.
      The equivalent in the EEF seems to be the appearance of the higher barrier, which makes it harder to ionize the second electron:
      Ignoring the time-dependence of the barrier in such a simulation, e.g.\ by using the TICA approximation, would overestimate the amount of electron density leaving the core region.
      
      The barrier is also a sign of excited states in the core region. 
      Looking at Fig.\ \ref{fig:es}, we see that the EEF potential of the first excited state $\psi_1$ has a higher barrier in the core region compared to that of the ground state and the corresponding average energy $\varepsilon_{\rm av}$ as well as the Fubini-Studi potential $\varepsilon_{\rm FS}$ look very similar to those shown for \unit[$F_0 = 0.050$]{$E\m{h} /(e a_0)$} in Fig. \ref{fig:potc1}.
      Occupation numbers $|c_j|^2$ with 
      \begin{align}
        c_j(t) = \braket{\psi_j(x_1,x_2)|\psi(x_1,x_2;t)}
        \label{eq:ccc}
      \end{align}
      confirm this observation, as they show significant population of lower excited states of our model Hamiltonian.
      An interesting case are the simulations for \unit[$\omega_0 = 0.2$]{$E_{\rm h}/\hbar$} where the laser frequency is very close to the transition between $\psi_0$ and $\psi_1$ (\unit[$E_1 - E_0 = 0.201$]{$E_{\rm h}$}).
      There, the initial potential $\varepsilon(x_1;t)$ resembles $\varepsilon_0(x_1)$ but becomes close to $\varepsilon_1(x_1)$ in the core region during the pulse, with a $t$-dependent variation that indicates some population of other states.
      
      \begin{figure}[htbp]
        \begin{center}
        \includegraphics[width=0.5\textwidth]{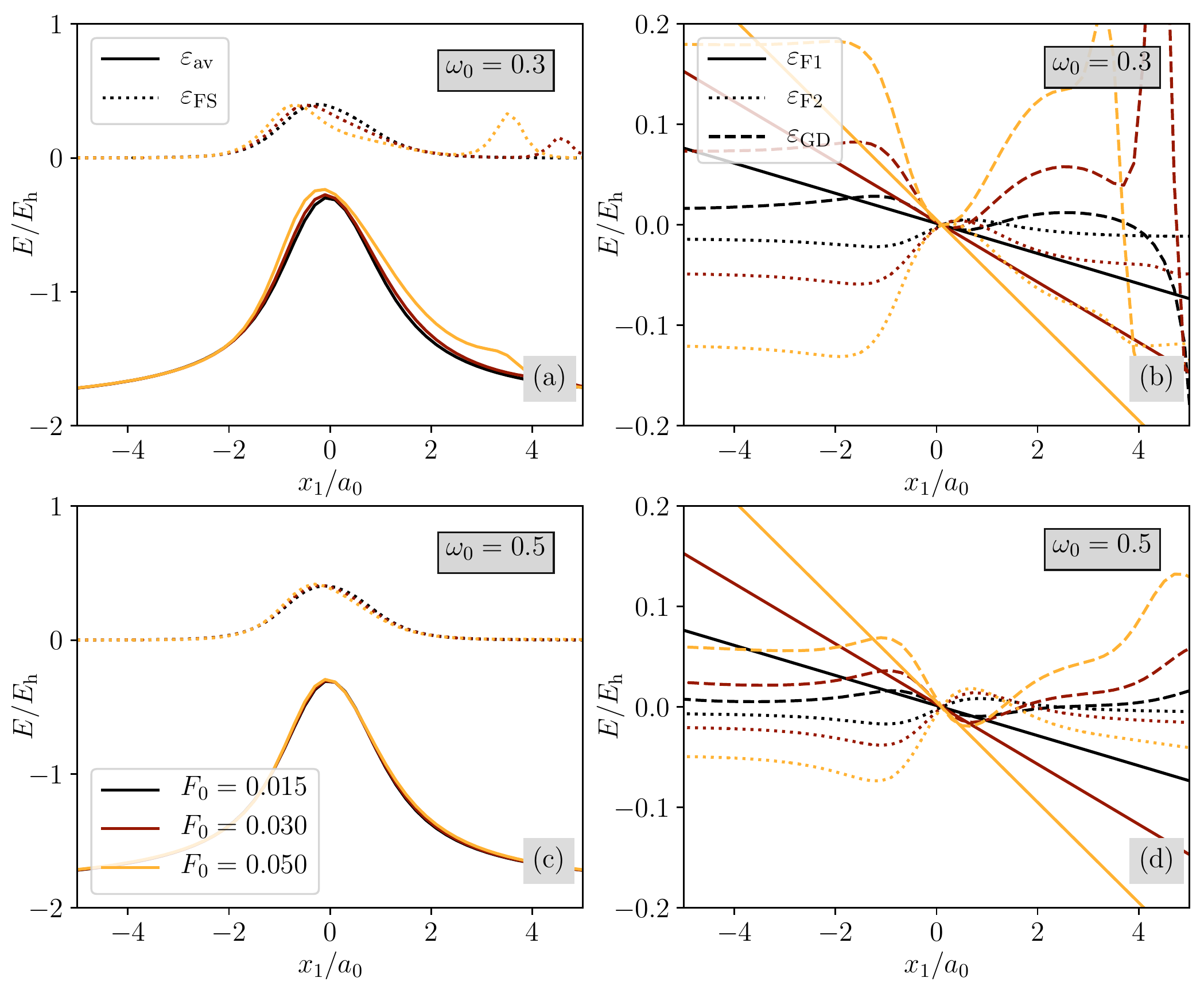}
        \caption{Like panels (d), (e) of Fig.\ \ref{fig:potc1}, but for laser frequencies \unit[$\omega_0=0.3$]{$E_{\rm h}/\hbar$} (top) and \unit[$\omega_0=0.5$]{$E_{\rm h}/\hbar$} (bottom).} \label{fig:potc}
        \end{center}
      \end{figure}
      
      Further information about the EEF potential $\varepsilon$ can be gained by looking at the one-electron laser interaction potential $\varepsilon_{\rm F1}$, the effective interaction potential $\varepsilon_{\rm F2}$ with the laser field, and the gauge-dependent part $\varepsilon_{\rm GD}$ of the EEF potential.
      For SAE calculations, recent publications have found that there is a screening effect due to polarization of the ``other'' electrons which cancels the effect of the laser potential in the core region \cite{romanov2020,abu-samha2020}.
      In the EEF, the behavior is somewhat different:
      First, we note that $\varepsilon_{\rm F2}$ and $\varepsilon_{\rm GD}$ cancel each other mostly, as illustrated in Fig.\ \ref{fig:potc1}, panels (e) and (f), for a laser frequency \unit[$\omega_0 = 0.1$]{$E_{\rm h}/\hbar$}.
      What remains is a potential well in the core region that is more pronounced with higher field strength $F_0$.
      It partially counteracts the one-electron laser interaction potential $\varepsilon_{\rm F1}$, as can be seen by comparing $\varepsilon_{\rm F1}$ and $\varepsilon_{\rm F2} + \varepsilon_{\rm GD}$ in the bottom-right panel of Fig.\ \ref{fig:potc1}.
      For smaller $F_0$ the effect of $\varepsilon_{\rm F2} + \varepsilon_{\rm GD}$ may indeed be approximated by ``switching off'' $\varepsilon_{\rm F1}$ in the core region, but for larger $F_0$ the potential well is relevant and modeling within the EEF framework seems to be more involved that what was proposed for SAE approaches.
      
      The situation is different for higher frequencies $\omega_0$, as illustrated in Fig. \ref{fig:potc}.
      We find that for higher $\omega_0$ both the average energy $\varepsilon_{\rm av}$ and the Fubini-Studi potential $\varepsilon_{\rm FS}$ have little $t$-dependence and can thus be approximated by the initial potentials.
      Also, the effective interaction potential $\varepsilon_{\rm F2}$ with the laser field and the gauge-dependent part of the potential $\varepsilon_{\rm GD}$ cancel almost perfectly in the core region, leaving only the one-electron laser interaction potential $\varepsilon_{\rm F1}$ as contribution to the total potential $\varepsilon$.
      
\begin{figure}[htbp]
        \begin{center}
        \includegraphics[scale=0.99]{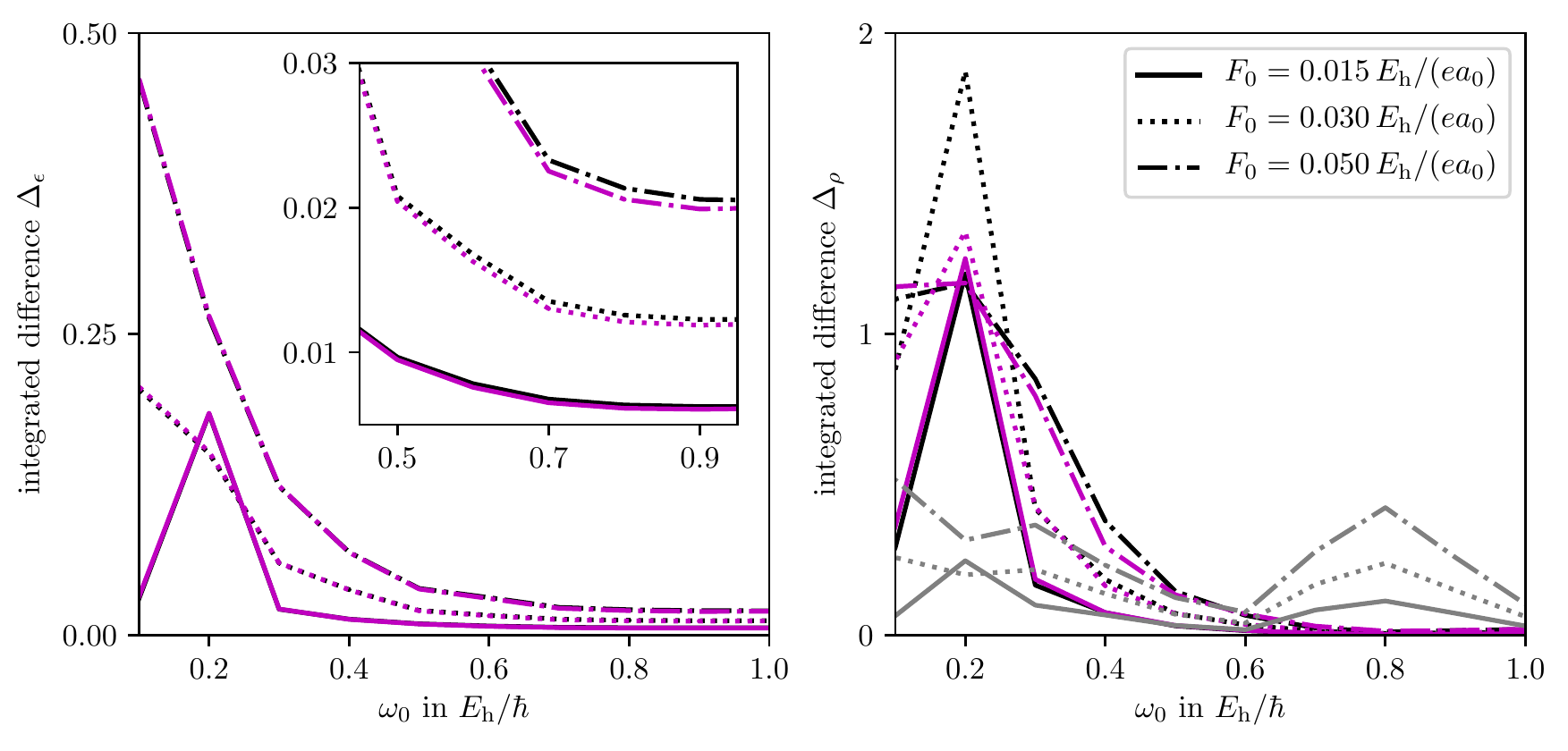}
        \caption{
        Left: Integrated absolute difference $\Delta_{\varepsilon}$ between the EEF potential and the TICA potential (black) as well as the TICA potential neglecting the modified dipole $d_0$ (magenta), for different field strengths $F_0$ and laser frequencies $\omega_0$.
        Right: Like left panel, but for the density difference $\Delta_{\rho}$. The gray lines show $\Delta_{\rho}$ for the SAE simulation.} \label{fig:ticac}
        \end{center}
      \end{figure}

      From those findings, we expect that a TICA simulation should be appropriate for high frequencies of the laser field because it is close to the EEF potential which represents the exact dynamics.
      To quantify this statement, we computed the integrated absolute difference 
      \begin{align}
        \Delta_{\varepsilon} = \frac{1}{6 T} \int\limits_0^{6T} \int |\varepsilon(x_1;t) - \varepsilon'(x_1;t) | \rho(x_1;t) \, dx_1 dt
        \label{eq:deleps}
      \end{align}
      between the EEF potential $\varepsilon(x_1;t)$ and the potential in a TICA simulation,
      \begin{align}
        \varepsilon'(x_1;t) = \varepsilon^{\rm TICA} + F(t) (x_1 + d_0(x_1)), \label{eq:tica01}
      \end{align}
      and without the dipole modification, 
      \begin{align}
        \varepsilon'(x_1;t) = \varepsilon^{\rm TICA} + F(t) x_1, \label{eq:tica02}
      \end{align}
      as well as the integrated density differences
      \begin{align}
        \Delta_{\rho} = \frac{1}{6 T} \int\limits_0^{6T} \int |\rho(x_1;t) - \rho'(x_1;t)| \, dx_1 dt
        \label{eq:delrho}
      \end{align}
      with the exact electron density $\rho(x_1;t)$ and with $\rho'(x_1;t)$ being either the density from an SAE simulation, the density from a TICA simulation using the potential \eqref{eq:tica01}, or the density from a TICA simulation without modified dipole using the potential \eqref{eq:tica02}.
      The absolute value of $\Delta_{\varepsilon}$ is weighted by the one-electron density $\rho(x_1;t)$ to count only relevant parts of the potential, and the time integration is performed over the duration $6 T$ of the 6-cycle laser pulse.
      As $T$ changes with the frequency, we also divide both differences by the pulse duration.
      The results are shown in Figure \ref{fig:ticac}.
      As expected, $\Delta_{\varepsilon}$ becomes smaller with increasing frequency $\omega_0$, hence the TICA potential becomes closer to the EEF potential.
      A notable exception is when \unit[$\omega_0 = 0.2$]{$E_{\rm h}/\hbar$}, which is close to resonance of the transition between the ground state and the first excited state:
      The TICA potential has, by construction, always the character of the ground state.
      In contrast, the time-dependent EEF potential resembles the ground state EEF potential only initially but becomes close to the EEF potential of the first excited state during the propagation.
      Neglect of $d_0$ always make the agreement better for $\Delta_{\varepsilon}$, but only slightly.
      However, it needs to be tested if $d_0$ plays a more prominent role when more electrons are part of the system.
      
      The density difference $\Delta_{\rho}$ illustrates that the different potentials influence the dynamics.
      Clearly, the agreement of the TICA densities (with and without modified dipole) becomes better with higher frequencies, while the SAE simulation is better than the TICA approximation for low frequencies but worse for high frequencies.
      A closer look at the dynamics shows what can be expected from the neglect of the time-dependent barrier in the TICA simulation for low frequencies:
      Far too much electron density leaves the core region and becomes highly delocalized.
      In contrast, the SAE simulation captures the dynamics qualitatively correctly for low frequencies.
      The applicability of the TICA approximation thus seems complementary to the SAE assumption, as the latter is often applied for relatively low frequencies $\omega_0$ (in the visible regime, e.g.\ for \unit[800]{nm} laser radiation) and is considered a good description of tunnel ionization in the strong field.
      
      We note that, interestingly, the ionization yield at the grid boundaries is well reproduced with the TICA simulations and is, for low frequencies, in even better agreement with the exact ionization yield than what is obtained from an SAE simulation.
      However, this finding is a coincidence for our model, because the dynamics of the TICA simulation differs drastically from the exact simulation for these low frequencies of the laser field.
      
      \begin{figure}[htbp]
        \begin{center}
        \includegraphics[scale=0.6]{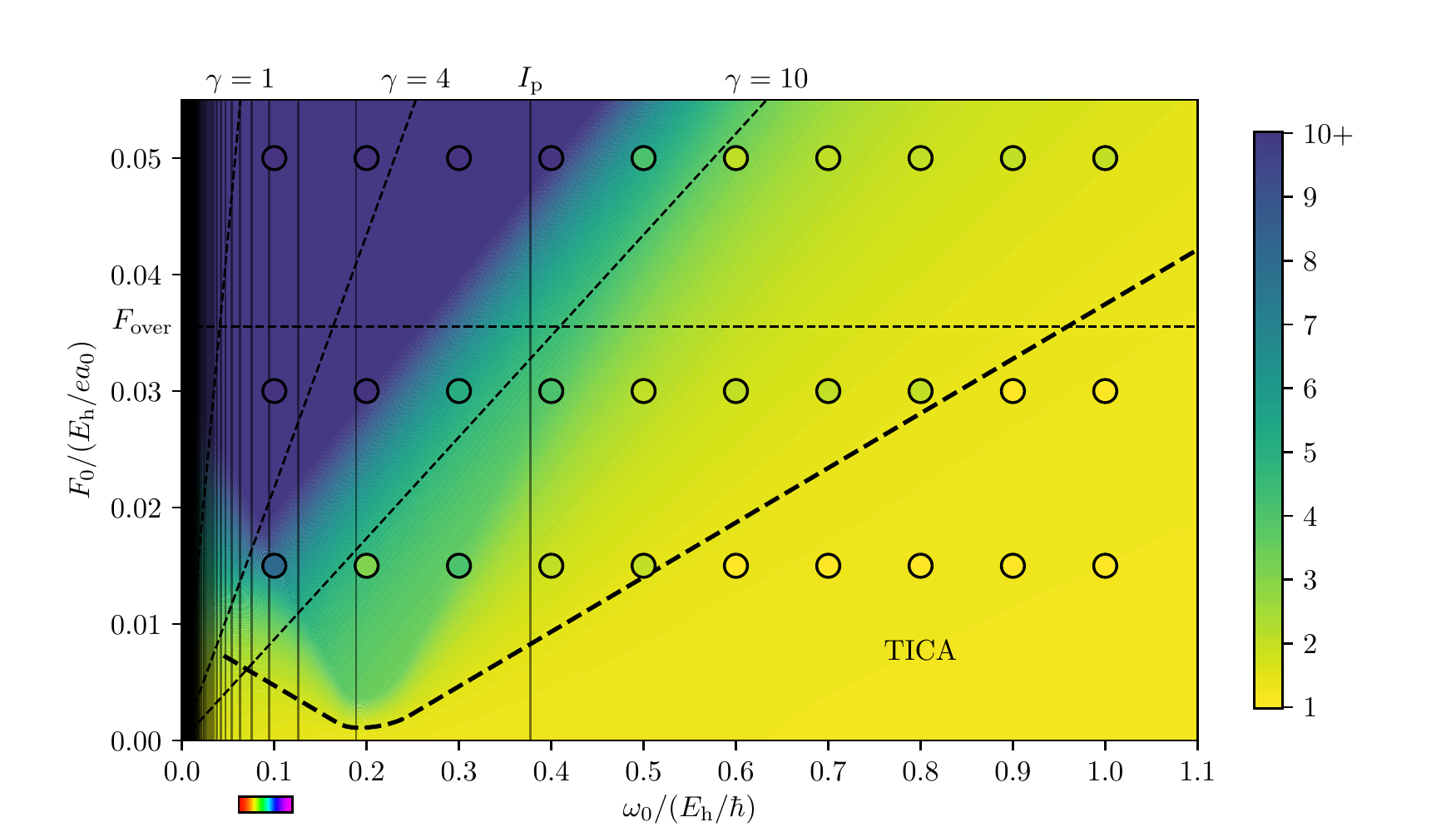}
        \caption{Like Fig.\ \ref{fig:para}, but with number of eigenstates needed to model the Exact Electron Factorization potential indicated as color/shade. The thick black dashed line indicates the approximate bound where the Time-Independent Conditional Amplitude approximation is valid for our model.} \label{fig:com}
        \end{center}
      \end{figure}
      
      The TICA approximation is based on one electronic state only.
      To understand better what is needed in the EEF framework to correctly describe the dynamics beyond the TICA approximation, we determined how many states $\psi_j$ are actually needed to reproduce the dynamics of the system.
      Based on the expansion coefficients \eqref{eq:ccc} of the bound states $\psi_j$ during a propagation, EEF potentials were constructed from the truncated wavefunction 
      \begin{align}
        \psi_{n_{\rm e}}(x_1,x_2;t) = \sum_{j=0}^{n_{\rm e}} c_j(t) \psi_j(x_1,x_2)
      \end{align}
      and compared to the exact potentials in the core region.
      Fig.\ \ref{fig:com} shows graphically the index $n_{\rm e}$ of the highest excited state needed to reasonably reproduce the EEF potential.
      Less states may be necessary as it may happen that some states with $j < n_{\rm e}$ are not populated.
      To find the highest state to be included, we also calculated the occupation numbers based on a Rabi model of the ground and the first excited state. 
      Starting with the initial occupation numbers $|c_0|^2 = 1$, $|c_1|^2 = 0$, within the rotating-wave approximation the occupation numbers evolve with $t$ as
      \begin{equation}
        \begin{pmatrix} |c_0 (t)|^2 \\ |c_1 (t)|^2 \end{pmatrix} = \begin{pmatrix}
        \frac{\delta^2+ |\Omega|^2 \cos^2(\omega\m{R} t)}{ \delta^2+|\Omega|^2} \\
        \frac{|\Omega|^2 \sin^2(\omega\m{R} t)}{ \delta^2+|\Omega|^2}   
        \end{pmatrix} \, ,
      \end{equation}
      where $\delta = E_1 - E_0 - \omega_0$, $\Omega = F_0 \braket{\psi_0|x_1+x_2 |\psi_1}$, and $\omega\m{R} = \frac{1}{2} \sqrt{\delta^2+|\Omega|^2}$. 
      Here, we consider the occupation numbers only until the end of the pulse.
      The thick dashed line in Fig.\ \ref{fig:com} shows where the transition from a one-state to a multi-state model approximately is located, based on the criterion that $|c_1 (t)|^2$ does not exceed \unit[0.8]{\%}.
      
      From Fig.\ \ref{fig:com} we see where TICA is expected to reproduce the dynamics accurately:
      For high laser field frequencies $\omega_0$ or for large Keldysh parameter $\gamma$, as well as for small $\omega_0$ and laser field strength $F_0$ (smaller than the field strengths $F_0$ used in our simulations), where the initial state is only little perturbed.
      For parameter regions around $\gamma = 1$ and the frequencies of visible light, which is where tunnel ionization happens and where a lot of activity in attoscience was focused in recent years, many states are needed to reproduce the exact EEF potential and thus the exact dynamics.
      
    \section{Conclusions}
      
      \label{sec:con}
      
      In the framework of the EEF, a many-electron dynamics can be mapped to a one-electron dynamics exactly.
      The effective potentials appearing in this one-electron dynamics carry a heavy burden, as they encode the time-dependent many-electron effects and the anti-symmetry requirements of the many-electron wavefunction.
      From the study of our simple model we found that to study ionization dynamics in laser fields, correct description of low-lying excitations in the core region is of central importance to obtain good effective potentials.
      However, we also found that some terms in the effective potential can be neglected.
      It will be interesting to study how the features found in the EEF potential carry over to more electrons.
      
      Additionally, we found that the simplest approximation of the EEF, the TICA approximation, provides a good description of the dynamics for relatively high frequencies of the laser field.
      It is thus complementary to the SAE assumption, which is typically used for comparably low frequencies.
      A TICA simulation for a realistic system is computationally as expensive as a SAE simulation, hence it is worthwhile to to to test how the TICA approach works compared to experimental data.
      However, the TICA approximation is based on only one electronic state of the many-electron system and does, for our model, not reproduce the exact dynamics for relatively low frequencies of the laser field.
      In this regime further electronic states are populated and a TICA simulation does not describe the core dynamics correctly. 
      
      Methods improving on the TICA approximation are conceivable, e.g.\ by simulating the dynamics of the bound states with an approach based e.g.\ on a few many-electron Slater determinants.
      Although there are problems due to a truncated dynamics in the core region \cite{ruggenthaler2009}, an approach based on that idea may provide suitable one-electron EEF potentials which reproduce the exact dynamics well.
      In this way, it would also be possible to avoid the difficult explicit coupling of bound and continuum states by simulating the bound-state dynamics and the ionization dynamics separately.
      Further developments are needed to find a practical method based on the EEF, but our analysis shows that the features of the exact EEF potentials can be understood and seem to be accessible from a computational point of view also for systems of experimental interest.

    \section{Acknowledgement}
    
      This research is supported by an Ambizione grant of the Swiss National Science Foundation (grant number 174212).
    
  \bibliographystyle{unsrt}
  \bibliography{eef}

\begin{thebibliography}{10}

\bibitem{pazourek2015}
Renate Pazourek, Stefan Nagele, and Joachim Burgd\"orfer.
\newblock Attosecond chronoscopy of photoemission.
\newblock {\em Rev. Mod. Phys.}, 87:765--802, Aug 2015.

\bibitem{calegari2016}
Francesca Calegari, Giuseppe Sansone, Salvatore Stagira, Caterina Vozzi, and
  Mauro Nisoli.
\newblock {Advances in attosecond science}.
\newblock {\em Journal of Physics B: Atomic, Molecular and Optical Physics},
  49(6):062001, feb 2016.

\bibitem{nisoli2017}
Mauro Nisoli, Piero Decleva, Francesca Calegari, Alicia Palacios, and Fernando
  Mart\'in.
\newblock Attosecond electron dynamics in molecules.
\newblock {\em Chemical Reviews}, 117(16):10760--10825, 2017.
\newblock PMID: 28488433.

\bibitem{biswas2020}
Shubhadeep Biswas, Benjamin F\"org, Lisa Ortmann, Johannes Sch\"otz, Wolfgang
  Schweinberger, Tom\'{a}\v{s} Zimmermann, Liangwen Pi, Denitsa Baykusheva,
  Hafiz~A. Masood, Ioannis Liontos, Amgad~M. Kamal, Nora~G. Kling, Abdullah~F.
  Alharbi, Meshaal Alharbi, Abdallah~M. Azzeer, Gregor Hartmann, Hans~J.
  W\"orner, Alexandra~S. Landsman, and Matthias~F. Kling.
\newblock {Probing molecular environment through photoemission delays}.
\newblock {\em Nature Physics}, 16:778, 2020.

\bibitem{palacios2019}
Alicia Palacios and Fernando Mart\'in.
\newblock The quantum chemistry of attosecond molecular science.
\newblock {\em WIREs Computational Molecular Science}, n/a(n/a):e1430, 2019.

\bibitem{bruner2017}
Adam Bruner, Samuel Hernandez, François Mauger, Paul~M. Abanador, Daniel~J.
  LaMaster, Mette~B. Gaarde, Kenneth~J. Schafer, and Kenneth Lopata.
\newblock {Attosecond Charge Migration with TDDFT: Accurate Dynamics from a
  Well-Defined Initial State}.
\newblock {\em The Journal of Physical Chemistry Letters}, 8(17):3991--3996,
  2017.
\newblock PMID: 28792225.

\bibitem{sato2018}
Shunsuke~A. Sato, Hannes H\"ubener, Angel Rubio, and Umberto De~Giovannini.
\newblock {First-principles simulations for attosecond photoelectron
  spectroscopy based on time-dependent density functional theory}.
\newblock {\em The European Physical Journal B}, 91:126, 2018.

\bibitem{majorosi2018}
Szil\'ard Majorosi, Mih\'aly~G. Benedict, and Attila Czirj\'ak.
\newblock Improved one-dimensional model potentials for strong-field
  simulations.
\newblock {\em Phys. Rev. A}, 98:023401, Aug 2018.

\bibitem{majorosi2020}
Szil\'ard Majorosi, Mih\'aly~G. Benedict, Ferenc Bog\'ar, G\'abor Paragi, and
  Attila Czirj\'ak.
\newblock {Density-based one-dimensional model potentials for strong-field
  simulations in He, ${\mathrm{H}}_{2}{}^{+}$, and ${\mathrm{H}}_{2}$}.
\newblock {\em Phys. Rev. A}, 101:023405, Feb 2020.

\bibitem{Schafer1993}
K.~J. Schafer, Baorui Yang, L.~F. DiMauro, and K.~C. Kulander.
\newblock Above threshold ionization beyond the high harmonic cutoff.
\newblock {\em Phys. Rev. Lett.}, 70:1599--1602, Mar 1993.

\bibitem{Yang1993}
Baorui Yang, K.~J. Schafer, B.~Walker, K.~C. Kulander, P.~Agostini, and L.~F.
  DiMauro.
\newblock Intensity-dependent scattering rings in high order above-threshold
  ionization.
\newblock {\em Phys. Rev. Lett.}, 71:3770--3773, Dec 1993.

\bibitem{Walker1994}
B.~Walker, B.~Sheehy, L.~F. DiMauro, P.~Agostini, K.~J. Schafer, and K.~C.
  Kulander.
\newblock Precision measurement of strong field double ionization of helium.
\newblock {\em Phys. Rev. Lett.}, 73:1227--1230, Aug 1994.

\bibitem{Awasthi2008}
Manohar Awasthi, Yulian~V. Vanne, Alejandro Saenz, Alberto Castro, and Piero
  Decleva.
\newblock Single-active-electron approximation for describing molecules in
  ultrashort laser pulses and its application to molecular hydrogen.
\newblock {\em Phys. Rev. A}, 77:063403, Jun 2008.

\bibitem{Ivanov2014}
I.~A. Ivanov and A.~S. Kheifets.
\newblock Strong-field ionization of {He} by elliptically polarized light in
  attoclock configuration.
\newblock {\em Phys. Rev. A}, 89:021402, Feb 2014.

\bibitem{abu-samha2010}
M.~Abu-samha and L.~B. Madsen.
\newblock {Single-active-electron potentials for molecules in intense laser
  fields}.
\newblock {\em Phys. Rev. A}, 81:033416, Mar 2010.

\bibitem{gordon2006}
Ariel Gordon, Franz~X. K\"artner, Nina Rohringer, and Robin Santra.
\newblock {Role of Many-Electron Dynamics in High Harmonic Generation}.
\newblock {\em Phys. Rev. Lett.}, 96:223902, Jun 2006.

\bibitem{Ishikawa2015}
K.~L. {Ishikawa} and T.~{Sato}.
\newblock A review on ab initio approaches for multielectron dynamics.
\newblock {\em IEEE Journal of Selected Topics in Quantum Electronics},
  21(5):1--16, Sep. 2015.

\bibitem{romanov2020}
A.~A. Romanov, A.~A. Silaev, M.~V. Frolov, and N.~V. Vvedenskii.
\newblock {Influence of the polarization of a multielectron atom in a strong
  laser field on high-order harmonic generation}.
\newblock {\em Phys. Rev. A}, 101:013435, Jan 2020.

\bibitem{abu-samha2020}
Mahmoud Abu-samha and Lars~Bojer Madsen.
\newblock {Effect of multielectron polarization in the strong-field ionization
  of the oriented CO molecule}.
\newblock {\em Phys. Rev. A}, 101:013433, Jan 2020.

\bibitem{abedi2010}
Ali Abedi, Neepa~T. Maitra, and E.~K.~U. Gross.
\newblock {Exact Factorization of the Time-Dependent Electron-Nuclear Wave
  Function}.
\newblock {\em Phys. Rev. Lett.}, 105:123002, Sep 2010.

\bibitem{abedi2012}
Ali Abedi, Neepa~T. Maitra, and E.~K.~U. Gross.
\newblock {Correlated electron-nuclear dynamics: Exact factorization of the
  molecular wavefunction}.
\newblock {\em J. Chem. Phys.}, 137:22A530, 2012.

\bibitem{gonze2018}
Xavier Gonze, Jianqiang~Sky Zhou, and Lucia Reining.
\newblock {Variations on the ``exact factorization'' theme}.
\newblock {\em The European Physical Journal B}, 91:224, 2018.

\bibitem{agostini2018}
Federica Agostini, Ivano Tavernelli, and Giovanni Ciccotti.
\newblock {Nuclear quantum effects in electronic (non)adiabatic dynamics}.
\newblock {\em The European Physical Journal B}, 91:139, 2018.

\bibitem{briggs2000}
J.~S. Briggs and J.~M. Rost.
\newblock {Time dependence in quantum mechanics}.
\newblock {\em The European Physical Journal D}, 10:311, 2000.

\bibitem{schild2018}
Axel Schild.
\newblock {Time in quantum mechanics: A fresh look at the continuity equation}.
\newblock {\em Phys. Rev. A}, 98:052113, Nov 2018.

\bibitem{cederbaum2015}
Lorenz~S. Cederbaum.
\newblock {The exact wavefunction of interacting N degrees of freedom as a
  product of N single-degree-of-freedom wavefunctions}.
\newblock {\em Chemical Physics}, 457:129 -- 132, 2015.

\bibitem{lacombe2020}
Lionel Lacombe and Neepa~T. Maitra.
\newblock Embedding via the exact factorization approach.
\newblock {\em Phys. Rev. Lett.}, 124:206401, May 2020.

\bibitem{schild2017}
Axel Schild and E.~K.~U. Gross.
\newblock {Exact Single-Electron Approach to the Dynamics of Molecules in
  Strong Laser Fields}.
\newblock {\em Phys. Rev. Lett.}, 118:163202, 2017.

\bibitem{hunter1986}
Geoffrey Hunter.
\newblock {The exact one‐electron model of molecular structure}.
\newblock {\em Int. J. Quant. Chem.}, 29:197, 1986.

\bibitem{hunter1987}
Geoffrey Hunter.
\newblock {The Exact Schr\"odinger Equation for the Electron Density}.
\newblock In R.~Erdahl and Jr. V.H.~Smith, editors, {\em Density Matrices and
  Density Functionals}, pages 583--596. D. Reidel Publishing Company,
  Dordrecht, Holland, 1987.

\bibitem{kraisler2020}
Eli Kraisler.
\newblock {Asymptotic Behavior of the Exchange-Correlation Energy Density and
  the Kohn-Sham Potential in Density Functional Theory: Exact Results and
  Strategy for Approximations}.
\newblock {\em Israel Journal of Chemistry}, n/a(n/a), 2020.

\bibitem{brabec2005}
Thomas Brabec, Michel C\^ot\'e, Paul Boulanger, and Lora Ramunno.
\newblock Theory of tunnel ionization in complex systems.
\newblock {\em Phys. Rev. Lett.}, 95:073001, Aug 2005.

\bibitem{zhao2007}
Zengxiu Zhao and Thomas Brabec.
\newblock Tunnel ionization in complex systems.
\newblock {\em Journal of Modern Optics}, 54(7):981--997, 2007.

\bibitem{shpilkin1996}
Sergey~A. Shpilkin, Evgenii~A. Smolenskii, and Nikolai~S. Zefirov.
\newblock Topological structure of the configuration space and the separation
  of spin and spatial variables for n-electron systems.
\newblock {\em Journal of Chemical Information and Computer Sciences},
  36(3):409--412, 1996.

\bibitem{provost1980}
J.~P. Provost and G.~Vallee.
\newblock {Riemannian structure on manifolds of quantum states}.
\newblock {\em Communications in Mathematical Physics}, 76:289, 1980.

\bibitem{berry1989}
Michael~V. Berry.
\newblock {The Quantum Phase, Five Years After}.
\newblock In A.Shapere and F.Wilczek, editors, {\em Geometric Phases in
  Physics}, page~7. World Scientific, 1989.

\bibitem{requist16}
Ryan Requist, Falk Tandetzky, and E.~K.~U. Gross.
\newblock {Molecular geometric phase from the exact electron-nuclear
  factorization}.
\newblock {\em Phys. Rev. A}, 93:042108, 2016.

\bibitem{Bauer1997}
D.~Bauer.
\newblock Two-dimensional, two-electron model atom in a laser pulse: Exact
  treatment, single-active-electron analysis, time-dependent density-functional
  theory, classical calculations, and nonsequential ionization.
\newblock {\em Phys. Rev. A}, 56:3028--3039, Oct 1997.

\bibitem{pazourek2012}
Renate Pazourek, Johannes Feist, Stefan Nagele, and Joachim Burgd\"orfer.
\newblock {Attosecond Streaking of Correlated Two-Electron Transitions in
  Helium}.
\newblock {\em Phys. Rev. Lett.}, 108:163001, Apr 2012.

\bibitem{finzel2016}
Kati Finzel.
\newblock Local conditions for the pauli potential in order to yield
  self-consistent electron densities exhibiting proper atomic shell structure.
\newblock {\em The Journal of Chemical Physics}, 144(3):034108, 2016.

\bibitem{Keldysh1965}
LV~Keldysh et~al.
\newblock Ionization in the field of a strong electromagnetic wave.
\newblock {\em Sov. Phys. JETP}, 20(5):1307--1314, 1965.

\bibitem{kiyan1991}
I.~Yu. Kiyan and V.P. Krainov.
\newblock {Above-barrier ionization of the hydrogen atom in a superstrong
  optical field}.
\newblock {\em Sov. Phys. JETP}, 73:429, 1991.

\bibitem{amini2019}
Kasra Amini, Jens Biegert, Francesca Calegari, Alexis Chac{\'{o}}n, Marcelo~F
  Ciappina, Alexandre Dauphin, Dmitry~K Efimov, Carla~Figueira
  de~Morisson~Faria, Krzysztof Giergiel, Piotr Gniewek, Alexandra~S Landsman,
  Micha{\l} Lesiuk, Micha{\l} Mandrysz, Andrew~S Maxwell, Robert
  Moszy{\'{n}}ski, Lisa Ortmann, Jose~Antonio P{\'{e}}rez-Hern{\'{a}}ndez,
  Antonio Pic{\'{o}}n, Emilio Pisanty, Jakub Prauzner-Bechcicki, Krzysztof
  Sacha, Noslen Su{\'{a}}rez, Amelle Zaïr, Jakub Zakrzewski, and Maciej
  Lewenstein.
\newblock Symphony on strong field approximation.
\newblock {\em Reports on Progress in Physics}, 82(11):116001, oct 2019.

\bibitem{qmstunfti}
{QMstunfti -- Quantum Dynamics with (Sparse) Matrix Representations in Python}.
\newblock \url{https://gitlab.com/axelschild/QMstunfti}.

\bibitem{scipy}
{SciPy}.
\newblock \url{https://scipy.org/}.

\bibitem{arpack}
R.~B. Lehoucq, D.~C. Sorensen, and C.~Yang.
\newblock {\em {ARPack User's Guide: Solution of Large-Scale Eigenvalue
  Problems}}.
\newblock SIAM, 1998.

\bibitem{Hofmann2014}
C.~Hofmann, A.~S. Landsman, A.~Zielinski, C.~Cirelli, T.~Zimmermann,
  A.~Scrinzi, and U.~Keller.
\newblock Interpreting electron-momentum distributions and nonadiabaticity in
  strong-field ionization.
\newblock {\em Phys. Rev. A}, 90:043406, Oct 2014.

\bibitem{elliott2012b}
Peter Elliott and Neepa~T. Maitra.
\newblock Propagation of initially excited states in time-dependent
  density-functional theory.
\newblock {\em Phys. Rev. A}, 85:052510, May 2012.

\bibitem{elliott2012}
P.~Elliott, J.~I. Fuks, A.~Rubio, and N.~T. Maitra.
\newblock {Universal Dynamical Steps in the Exact Time-Dependent
  Exchange-Correlation Potential}.
\newblock {\em Phys. Rev. Lett.}, 109:266404, Dec 2012.

\bibitem{hodgson2017}
M.~J.~P. Hodgson, Eli Kraisler, Axel Schild, and E.~K.~U. Gross.
\newblock {How Interatomic Steps in the Exact Kohn–Sham Potential Relate to
  Derivative Discontinuities of the Energy}.
\newblock {\em The Journal of Physical Chemistry Letters}, 8(24):5974--5980,
  2017.
\newblock PMID: 29179553.

\bibitem{kraisler2020b}
Eli Kraisler and Axel Schild.
\newblock {Discontinuous behavior of the Pauli potential in density functional
  theory as a function of the electron number}.
\newblock {\em Phys. Rev. Research}, 2:013159, Feb 2020.

\bibitem{lein2005}
Manfred Lein and Stephan K\"ummel.
\newblock {Exact Time-Dependent Exchange-Correlation Potentials for
  Strong-Field Electron Dynamics}.
\newblock {\em Phys. Rev. Lett.}, 94:143003, Apr 2005.

\bibitem{ruggenthaler2009}
M.~Ruggenthaler and D.~Bauer.
\newblock Rabi oscillations and few-level approximations in time-dependent
  density functional theory.
\newblock {\em Phys. Rev. Lett.}, 102:233001, Jun 2009.

\end{thebibliography}

\end{document}